\documentclass[12pt]{article}
\usepackage{indentfirst,psfig,times}
\usepackage{apalike}

\title{
Modulation of Base Specific Mutation and Recombination Rates
Enables Functional Adaptation within
the Context of the Genetic Code
}
\author{
Taison Tan$^1$, Leonard D.\ Bogarad$^2$, Michael W.\ Deem$^1$\\
~\\
~\\
\hbox{}$^1$Department of Bioengineering and
Department of Physics \& Astronomy\\
Rice University\\
Houston, TX 77005--1892\\
~\\
\hbox{}$^2$Division of Biology\\
California Institute of Technology\\
Pasadena, CA  91125
}

\begin{document}
\maketitle

\renewcommand{\baselinestretch}{1.3} \tiny\normalsize

{\flushleft
Corresponding author: Michael W. Deem, Rice University,
6100 Main Street---MS 142,
Houston, TX  77005-1892, 713-348-5852, 
713-348-5811 (fax), mwdeem@rice.edu.
\bigskip

}

\newpage

\centerline{\textbf{ABSTRACT}}
\begin{quote}

The persistence of life requires populations to adapt at a rate 
commensurate with the dynamics of their environment. Successful populations
that inhabit highly variable environments have evolved mechanisms to increase
the likelihood of successful adaptation.
%
We introduce a $64 \times 64$ matrix
to quantify base-specific mutation potential, analyzing  four different
replicative systems, error-prone PCR, mouse
antibodies, a nematode, and \emph{Drosophila}.
Mutational tendencies are correlated with the structural
evolution of proteins.
%
In systems under strong selective pressure,
mutational biases are shown to favor the
adaptive search of space, either by base mutation or by recombination.
Such adaptability is discussed within the context of the genetic
code at the levels of replication and codon usage.
\end{quote}
\newpage

\section{Introduction}

Viable mutations can potentiate the emergence of new life forms
and the adaptation of living organisms to new environmental
constraints.
Evolution occurs through a hierarchy of genetic events, including
base substitution, homologous recombination,
insertions, deletions, rearrangements, transpositions,  and
horizontal transfers \cite{Lawrence1997,Pennisi1998}.
Systems such as the adaptive immune system use somatic hypermutation to
rapidly search protein space to combat infectious agents. Likewise,
error-prone PCR is used in molecular evolution
protocols to search space in order to optimize protein
function.
In addition,
pathogens and cancers have evolved effective dynamic mechanisms,
often predicated on base substitution, to
evade immune and therapeutic selection.
In HIV, for example, the high rate of viral mutation makes
the development of a vaccine difficult and results in the
rapid onset of resistance to many current drugs.
Indeed, there is a correspondence between 
the ability of HIV to evolve drug resistance, the drug regimen
given, and the genetic makeup of the strains present in a
patient \cite{Lathrop99}.
Crucial to a thorough understanding of the base substitution
process is a mathematically precise quantification of the various 
mutation rates.

The mutational machinery
of hypermutation and recombination is under environment-dependent
regulation \cite{Bull2000}.  Studies have shown that
regulation is possible both in the process of replication and
error correction \cite{Walker2000} and in the type of
polymerases expressed \cite{Friedberg2000,Storb2001}.
The mechanism for 
maintenance of adaptability traits is population-based and requires a 
dynamic environment.
That evolvability is a group selectable trait has been 
shown in simulations of digital organisms
\cite{Travis,Pepper,Ofria1999,Ray1997,Wagner1996,Altenberg}.
Many of the biochemical events necessary to modify adaptability
are known.  At the simplest level,
mutation of a single amino acid site in the \emph{Taq} Pol I
enzyme is sufficient to greatly modulate the accuracy of
DNA replication \cite{Patel2001}.

The goal of this paper is to show 
how species can use evolution in populations as a space
searching advantage in the context of the genetic code.
Base-to-base rates of
synonymous, conservative, and
non-conservative mutation tendencies for each codon are described,
thus allowing for the quantification of evolutionary potential.
Base-specific mutation rates are dependent upon the fidelity of the
replication machinery, flanking sequences, and other environmental conditions.
The base substitution rate is non-uniform because transitions are
typically
greatly favored over transversions, and purines are 
typically substituted at a greater
frequency than pyrimidines. However, different replication systems have
different base specific mutation probabilities. 
It is here argued that within the 
context of genetic code, the emergence of replication variants that modulate
not only the overall rate but also base-specific mutation rates
allows populations
to increase the probability of searching productive survival space under
dynamic environmental constraints. 
Our theory complements previous observations for the immune
system \cite{Kepler1997},  recent observations of
codon bias toward increased adaptability in Influenza A \cite{Plotkin2003},
and recent work on digital organisms showing how adaptability evolves
within a population
\cite{Travis,Pepper,Ofria1999,Ray1997,Wagner1996,Altenberg}.

A codon mutation matrix that defines in a precise manner the 
probability per round of a codon mutating by base substitution into
another codon is introduced.
 This matrix provides the rates of all possible
$64 \times 64$ mutations.  From these detailed rates, 
properties of the codons themselves can be calculated.
For example, 
the codon mutation matrix allows the classification of codons according
to their synonymous, conservative, or non-conservative mutabilities.
Codons that tend to mutate in a 
more dramatic, non-conservative fashion are characterized as having 
a higher evolutionary potential, allowing for a more
rapid short-term adaptation.

To describe mutabilities of codons, there currently exists the $K_{\rm S}$
and $K_{\rm A}$ notation \cite{LiWuLuo}. The parameter 
$K_{\rm S}$ describes the number of
synonymous substitutions per site, and $K_{\rm A}$ describes
the number of non-synonymous substitution per 
site. Because of its average character, and because it is
based upon sequences that have undergone selection, the
$K_{\rm S}$ and $K_{\rm A}$ description is limited to estimating the
number of synonymous and non-synonymous nucleotide substitutions
between exons of homologous genes.

Some approaches that exist are indirect measures of intrinsic adaptability
at the genetic level.  The PAM and BLOSUM matrices, for example,
describe mutabilities between amino acids
rather than between codons \cite{PAM,BLOSUM,Durbin}.
Moreover, a matrix of pure mutational tendencies is ideally 
constructed from data gathered from non-selected
genomic data, such as intron regions or pseudo-genes.
Yang and Kumar have developed what is known as the Q
matrix \cite{Yang}. This matrix quantifies the underlying
mutational pattern of nucleotide substitution. 
This $4 \times 4$ matrix, which deals with bases rather than codons,
will be useful in our development.
A codon mutation matrix based upon the assumption that
the ratio of transition to transversion mutation rates is constant
and that the ratio of nonsynonymous to synonymous mutation rates
is constant has been developed
\cite{Yang2}.  This matrix can capture mutational
data that are consistent with the assumption of equal
transition to transversion and nonsynonymous to synonymous
mutation rates.
Our $64 \times 64$ matrix separates the species-specific mutation
probabilities, and  it additionally allows us to quantify the efficacy,
type, and biases of subsequent codon mutation changes in the context
of the genetic code.

In the context of pathogen and disease evolution, the mutation matrix
can be a valuable tool to quantify mutation probabilities and to enable
the design of therapeutics and vaccines that would most effectively
target disease epitopes that have the lowest chance of evolutionary escape
 \cite{Freire2002}.
In the context of
laboratory evolution of proteins, or protein molecular
evolution \cite{Stemmer1997,Benkovic2000,Arnold2000}, 
knowing the tendencies of codons
to mutate synonymously, conservatively, or non-conservatively would 
be helpful in experiment design.

\section{Methods}
\subsection{The Codon Mutation Matrix}

As an approximation, it is initially assumed
 that each base in a codon mutates independently.
This allows the $64 \times 64$ codon mutation matrix  to be constructed
from the $4 \times 4$  base mutation matrix.  In particular
\begin{equation}
T_{ij} = t_{i_1 j_1} t_{i_2 j_2} t_{i_3 j_3} \ ,
\label{1}
\end{equation}
where $i$ is the number of the codon that will be mutated and $j$ is the number of the codon that results after the mutation,
with $1 \le i,j \le 64$.  The codon is denoted by $i_1 i_2 i_3$, where
$i_1$ is the first base in codon $i$,
$i_2$ is the second base, and $i_3$ is the third base,
with $1 \le i_1, i_2, i_3 \le 4$.  Similarly,
$j_1 j_2 j_3$ is the base triplet for codon $j$.
The probability of a mutation from codon $i$ to
codon $j$ in one round of replication is given by
the codon mutation matrix $T_{ij}$.  In this mathematical
representation, the probability per round of no mutation is given by
$T_{ii}$.  Since either a mutation occurs or no mutation
occurs, this matrix satisfies the constraint
\begin{equation}
\sum_{j=1}^{64} T_{ij} = 1 ~~~~~\rm{for~all~}i \ .
\label{2}
\end{equation}
The probability per round of a mutation from one base to another
is given by the base substitution matrix $t$.  
The base mutation matrix also satisfies conservation
of probability $\sum_{j_1=1}^{4} t_{i_1 j_1} = 1$.
This definition leads to what is known mathematically
as a discrete-time Markov process.
The base mutation matrix $t$ can be
constructed from information about the mutation frequency for the four
bases, A, C, G, and T.  
The non-diagonal elements of $t$ are 
derived from the 12 different independent rates of
mutation.  Typically the non-diagonal elements are small,
since the rate of mutation is on the order of $10^{-2}-10^{-6}$
per base per replication.
The diagonal elements of the base mutation matrix are computed from the
conservation of probability constraint.
The $64 \times 64$ codon mutation matrix is then constructed from the
$4 \times 4$ base mutation matrix by equation \ref{1}.
 Each element of
the $64 \times 64$ matrix thus gives the probability per round of one codon
mutating to another codon. One round, or codon mutation step, can
include zero, one, two, or three simultaneous base mutations.

The assumption that DNA bases mutate independently 
can be refined in the presence of
additional experimental data.  It is known, for example, that
flanking bases affect the base mutation rate in the hypervariable
region of mouse
antibodies \cite{Smith}.  Overall mutation rates have been measured for
base triplets, and this information can be used to refine the
codon mutation matrix.  If $\omega_i$ is the 
observed mutation rate for codon $i$,  the improved 
codon mutation matrix $T'$ is defined as
\begin{equation}
T'_{ij} = \frac{\omega_i}{z} T_{ij}  \ ,
\label{2a}
\end{equation}
where $z$ is a constant chosen so that the average mutation rate of
the codons remains unchanged by this operation:
$z = \sum_{i \ne j} w_i T_{ij} / \sum_{i \ne j} T_{ij}$.
Alternatively, the assumption of equal transition to transversion
and synonymous to nonsynonymous mutation rates may be
used to generate a refined codon mutation matrix \cite{Yang2},
although this will not be done in the present work.

The codon mutation matrix differs from organism to organism and is
constructed here for several specific systems.
Since comparative trends are of interest,
 the overall average base mutation rate is set to be the same in all species,
 $2 \times 10^{-5}$ per replication. 
A different average mutation rate for each species would simply adjust the
 overall scale of the codon mutation matrix.
In each case,  the
base mutation matrix is first constructed from available data, and then
equation \ref{1} is used to construct the full codon mutation
matrix. 

The $64 \times 64$ codon mutation
 matrix contains a total of 4096 elements, each
element calculated from Equation \ref{1} or Equation \ref{2a}.
 For each codon 
a synonymous, conservative, and
non-conservative mutability is defined.
  The synonymous mutability, for example,
is the sum of all of the elements
of the codon mutation matrix that 
change a codon by a
synonymous mutation.
Similarly, the conservative mutability is the
sum of all of the elements
of the codon mutation matrix that 
change a codon by a
conservative mutation.
A conservative mutation
occurs when a codon mutates to a codon that codes for a different
amino acid that is, however, similar to the amino acid originally
encoded.
  Amino acids are similar if they are in the same
group, and there are seven groups: neutral and polar,
positive and polar, negative and polar, nonpolar with ring, nonpolar
without ring, cysteine, and stop. Substitutions that change the amino
acid to a different group are defined as non-conservative, and
substitutions that retain the encoded amino acid are
defined as synonymous.
Finally, the non-conservative mutability is the
sum of all of the elements
of the codon mutation matrix that 
change a codon by a
non-conservative mutation.
These three mutability
values express the probability that a specific codon will mutate
synonymously, conservatively, or non-conservatively in one
round of  replication.

\subsection{Systems Studied}

The mutation frequencies of the \emph{Taq} polymerase
in error-prone PCR
are available 
and can be
extracted \cite{Maranas}.
In the context of protein molecular evolution
\cite{Stemmer1997,Benkovic2000,Arnold2000},
understanding the mutational process in error-prone PCR
is especially important.
The base mutation matrix for this, and the other systems,
is available in the Supplementary Information.
  The three
mutability values for each codon for this system
are shown in Figure \ref{figmastercodon}a.

The codon mutation matrix is also constructed for mutations in the
intronic V regions of mouse antibodies \cite{Smith}.
 Equation \ref{2a} is used to account
for the effect of flanking bases in the mutation process, using
JH/J$\kappa$ intronic data \cite{Shapiro}.
The mutability values for this system
are shown in Figure \ref{figmastercodon}b.


The data from non-long terminal repeat retrotransposable elements
are used to construct the $4 \times 4$ base mutation matrix for
\emph{Drosophila} \cite{Petrov}.  Only the data from the terminal
branches, representing ``dead-on-arrival,'' nonfunctional copies
that are unconstrained by selection were used.  These copies
evolve as pseudo-genes.

The last system for which a codon mutation matrix is constructed is
mitochondrial DNA from \emph{Haemonchus contortus} \cite{Nematode}.
This is a nematode
in the same subclass Rhabditia as \emph{Caenorhabditis elegans}.
Coding regions of mtDNA were used to allow for comparison with
codon usage data available in the literature. 
The base mutation matrix obtained from this data is treated as applicable
to nuclear DNA, and so the standard genetic code is used.
While use of intronic data from \emph{C.\ elegans} 
would be preferable, such data are difficult
to collect due to the extensive divergence between
\emph{C.\ elegans} 
and its near relative, \emph{C. briggsae}
(T.\ Blumenthal, personal communication, 2001).
The mutation rate data estimated by the mtDNA mutation rates
will not play an essential role in the analysis.


\subsection{The No-Bias Codon Mutabilities}

We are looking for biases in the underlying mutation rates of the
replication machinery, not for biases in the genetic code itself.  The
genetic code biases--that hydrophobic residues tend to mutate
to hydrophobic residues and that
hydrophilic residues tend to mutate to hydrophilic residues--are
well known \cite{Woese,Epstein,Goldberg,Fitch,Volkenstein}.
To investigate biases other than those induced by the genetic code,
a refinement to the codon mutability plots is made.
 This refinement
subtracts from each mutability a value termed as the ``no-bias''
value. The ``no-bias'' value comes from a $64 \times 64$ matrix that
is created by using a $4 \times 4$ matrix where each non-diagonal term
has equal mutation frequencies, \emph{e.g.}\ 
equal  transition and transversion rates.
In other words, the no-bias plots indicate which empirically derived
mutabilities are above or below those expected if all base substitutions
were equally likely.
This matrix serves as a baseline for unbiased mutation rates
within the context of the genetic code.
This no-bias transformation is not a correction: it is a refined
way to do the analysis.
The overall mutation rate of the no-bias codon mutation matrix
is made to be same as that of the original codon mutation matrix.
Synonymous, conservative, and non-conservative mutabilities
are calculated from this baseline $64 \times 64$ matrix and subtracted 
from the original mutabilities, Figure \ref{fignobias}.

\section{Results}

\subsection{Modulation of Codon Mutation Rates}

\begin{figure}[htbp]
\centering
\leavevmode
\psfig{file=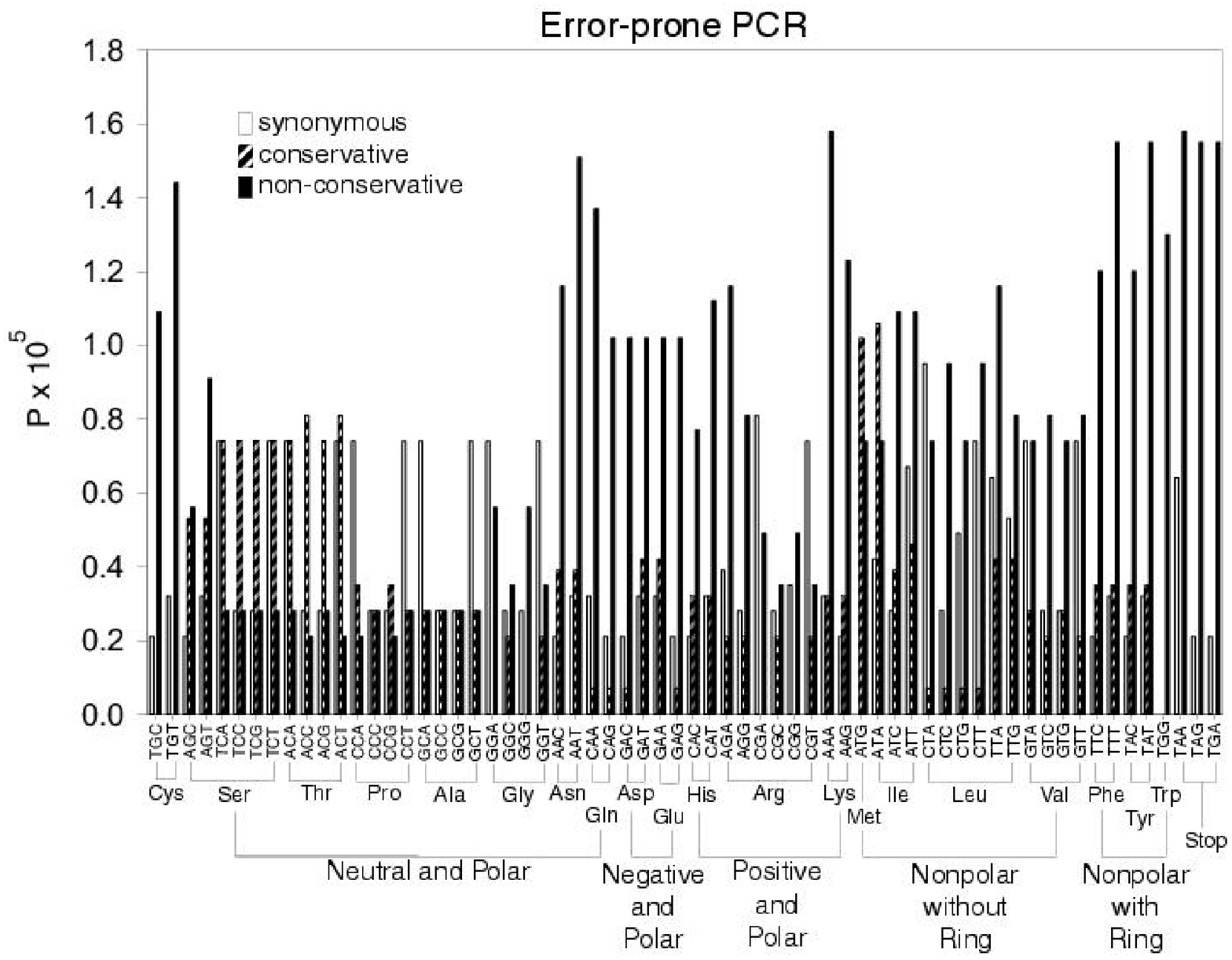,height=8cm}~\\
\centering
\leavevmode
\psfig{file=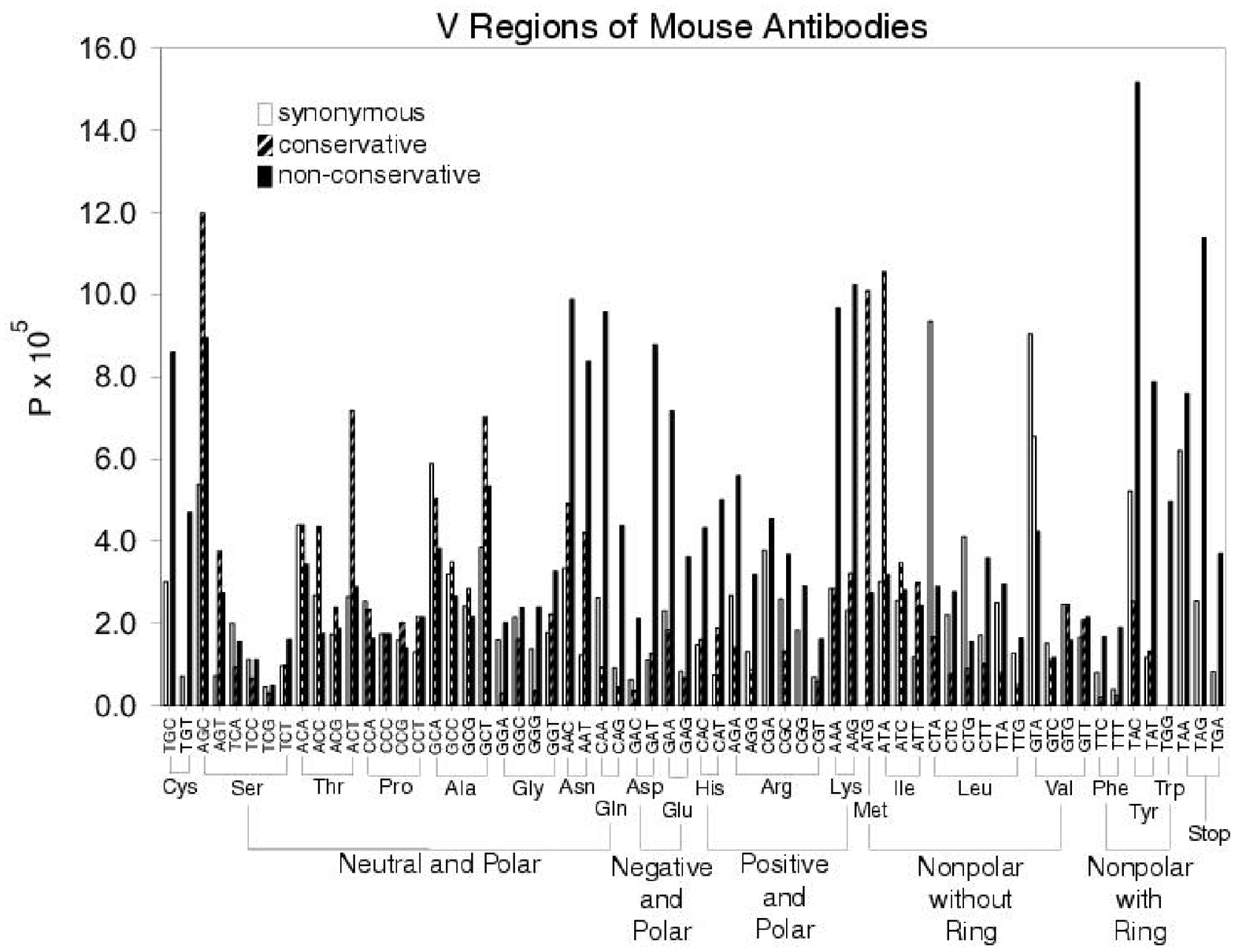,height=8cm}~\\
\caption{
The codon mutability plot for a) error-prone PCR
and b) V regions of mouse antibodies.
Each plot
displays the synonymous, conservative, and non-conservative
mutabilities for each codon.
}
\label{figmastercodon}
\end{figure}

\begin{figure}[htbp]
\centering
\leavevmode
\psfig{file=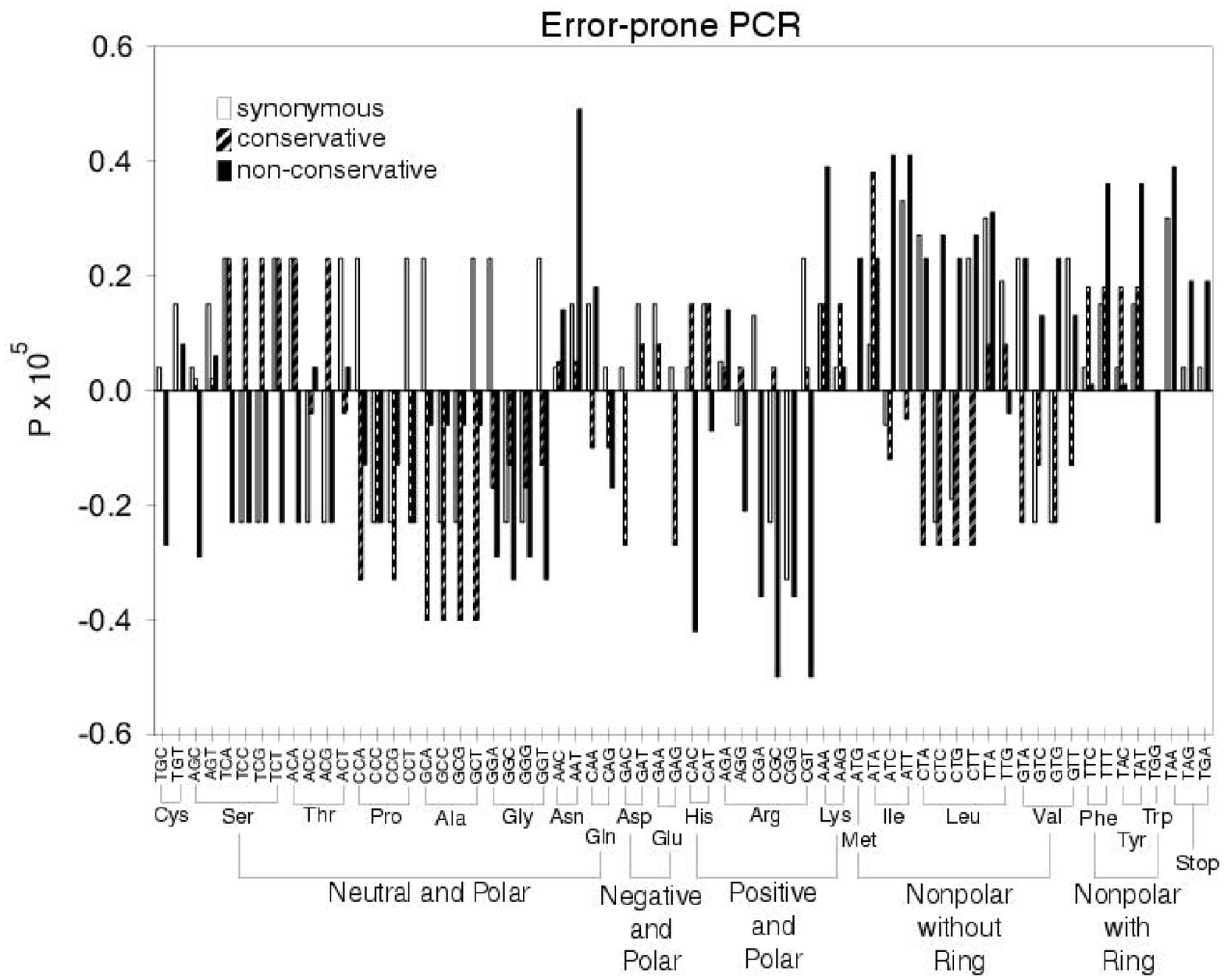,height=8cm}~\\
\centering
\leavevmode
\psfig{file=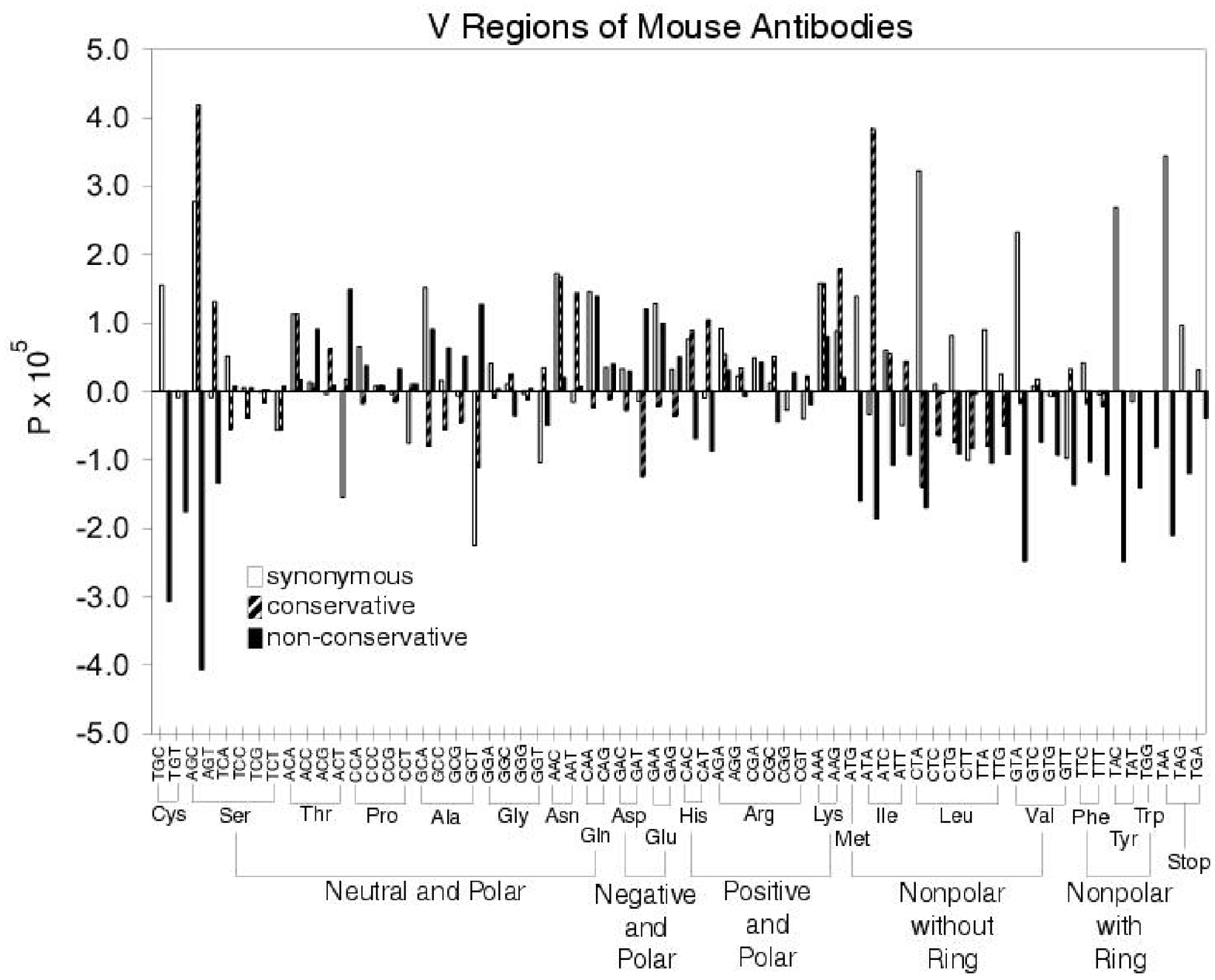,height=8cm}~\\
\caption{
The no-bias plot for a) error-prone PCR and
b) V regions of mouse antibodies.
This refinement to the codon mutability plots
takes into account the baseline substitution rate due to the
inherent structure in the genetic code.
}
\label{fignobias}
\end{figure}

Error-prone PCR, while not a pure biological system, is a central
tool and serves as an excellent example of the power of our approach.
Figure \ref{fignobias}a
 immediately reveals that for error-prone PCR, the codons that
code for polar amino
acids have low relative conservative and non-conservative
mutabilities. That is, these mutabilities are much lower than what
would be expected under unbiased conditions.
For the codons that code for the nonpolar amino acids, on
the other hand, a different
pattern is observed. In this case, the conservative and
non-conservative mutabilities are higher than the baseline values
generated from equal mutation rates.
Note that because of the factorization in Equation \ref{1},
our theory describes the biasing effect of base mutations,
and the ``reading frame'' of \emph{Taq} does not matter.
In Figures \ref{figmastercodon}a and \ref{fignobias}a we are showing the
effect of these biased base mutations when the ribosome
reads the exons in frame.

To study the possible effects of mutability modulation in a natural
population undergoing rapid, active evolution, the mouse V
regions are examined with the $64 \times 64$
 mutation matrix approach. Interestingly, higher conservative
and non-conservative
mutabilities are observed for the polar
amino acids compared to the
nonpolar amino acids, Figure \ref{fignobias}b.
We quantify the statistical significance of these results
by computing the probability per round that a random base mutation matrix
would lead to a ratio of mutation rates between the polar 
groups and the nonpolar groups that is as great or greater than that
observed.  That is, we take the ratio of the sum of the 
conservative and non-conservative
mutabilities from  Figure \ref{figmastercodon}
for these two groups.  
The probability by chance that this ratio is as large or larger than
that in Figure \ref{figmastercodon}b is 8.6\%.
From this extremely conservative statistic, it can be 
concluded that the pattern of increased mutability of 
polar amino acids is statistically significant to the level of 91\%.
We also perform this same calculation using another, independent
estimate of the base mutation matrix for mouse V regions \cite{Milstein1995}.
The probability by chance that the ratio of
conservative and non-conservative mutabilities
for a random matrix is larger than that given by this new matrix
is 5.3\%.  This result is, thus, significant to the level of 95\%.
It is interesting to note that if one assumed the experimentally measured
base mutation matrices were random,
\emph{i.e.}\ dominated by experimental noise, the probability that
two random such matrices would give a ratio as large or larger than
that observed in Figure \ref{figmastercodon}b is $0.086^2 = 0.7$\%.

It is difficult to measure experimentally
exact mutation rates. Thus, the sensitivity of the codon
mutation matrix to changes in the base substitution rates is of
interest.  In order to test the robustness of our findings for \emph{Taq} to
experimental noise, a random number is
added or subtracted from each of the twelve off-diagonal, independent
values in the $4 \times 4$ base mutation matrix.
This random number is generated from a Gaussian distribution with
zero mean and a standard deviation
that is equal to a given percentage of the average mutation
rate. This procedure generates a new $4 \times 4$ base mutation matrix,
from which a new $64 \times 64$ codon mutation matrix
is calculated.
To determine if the mutability bias patterns found in Figure
\ref{fignobias}a is perturbed by the addition of noise,
codon mutability plots are created with the new codon mutation matrix.
This plot displays the pattern observed in Figure \ref{fignobias}a
until the noise overwhelms the signal.
The pattern from Figure \ref{fignobias}a is still evident 
up to noise levels of 50\% of the average mutation rate, disappearing only
when the noise reaches 60\% \cite{Tan}.  An analagous calculation
was performed for the mouse V region system, and again the
pattern in Figure \ref{fignobias}b persisted up to noise levels of 50\% of the
average mutation rate, disappearing only
when the noise reaches 60\% \cite{Tan}.  
Thus, the observed trends in Figure \ref{fignobias} are rather robust
to the presence of experimental noise.

\begin{figure}[htbp]
\centering
\leavevmode
\psfig{file=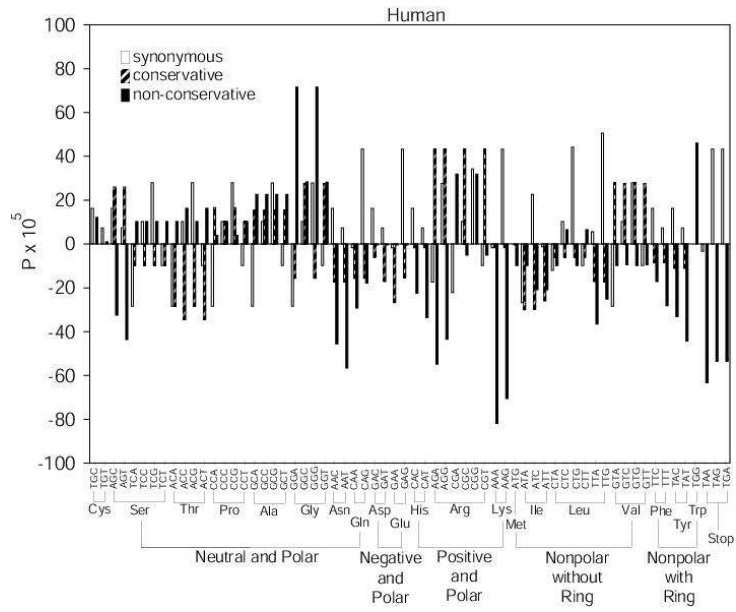,height=8cm}
\caption{No-bias plot for the non-immune-system genes
c-Myc, survivin1, survivin2, and TBP
in human B cells.}
\label{humanBcell}
\end{figure}

One might wonder whether this pattern of increased non-synonymous
mutabilities of charged residues would survive in other mouse or
mammalian genes.  Figure \ref{humanBcell} shows the no-bias codon
mutability plot derived from non-immune-system gene mutation rates from human
B cells.  Data are from \cite{Storb2000}.
As expected, there is no overall pattern.
A quantitative comparison to the polar to nonpolar ratio of
conservative and non-conservative mutabilities calculated for
Figure \ref{figmastercodon}b shows that in this case
the probability that a random
base mutation matrix has a value higher than that observed in
Figure \ref{humanBcell} is 25\%.
Thus, the increase in the non-synonymous mutability of the mammalian,
immunoglubulin V region in Figure \ref{figmastercodon}b
is unique and statistically significant.

\begin{figure}[htbp]
\centering
\leavevmode
\psfig{file=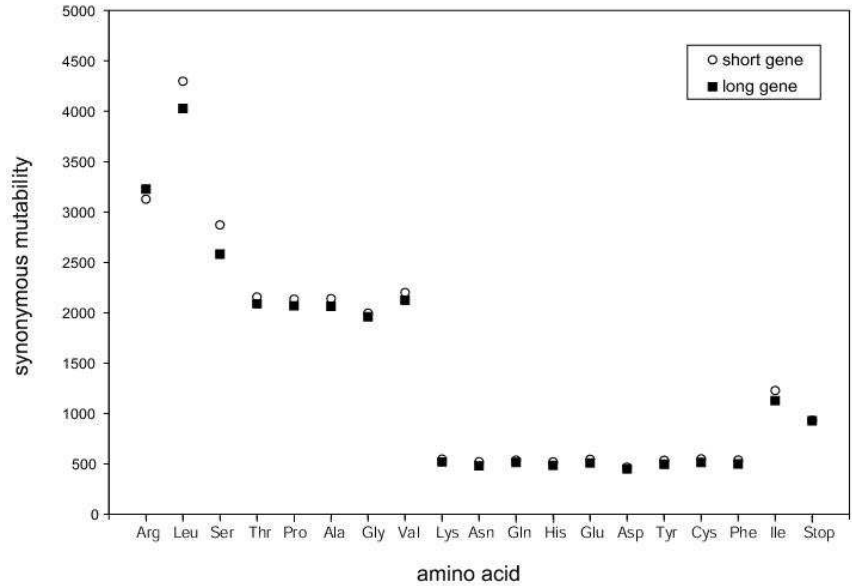,height=8cm}
\caption{Synonymous mutabilities for \emph{Drosophila}
for amino acids in short ($< 333$ amino acids) 
and long ($> 570$ amino acids) genes at high expression levels
(top 1/3 of genes with non-zero EST abundance).
Higher values of synonymous mutability are observed in the shorter genes.}
\label{drosynon}
\end{figure}

Further analysis of the codon mutation matrices was done by combining
mutability information with codon usage information.
Codon usage is necessary to determine via the mutation matrix
the average rate of mutation of a
gene, since the total rate of mutation depends on both the mutation
rate per codon and on which codons are present. 
By summing the product of the RSCU value \cite{Sharp1986}
and the
synonymous mutability for all the codons that code for a given
amino acid, the synonymous mutability of amino acid $\alpha$
is calculated:
\begin{equation}
{\rm synonymous~mutability} (\alpha) = \sum_{i \in \alpha}
p_i^\alpha \times {\rm synonymous~mutability} (i) \ , 
\end{equation}
where the synonymous mutability of codon $i$ is taken from
Figures  \ref{figmastercodon}a--\ref{figmastercodon}c,
and the codon usage $p_i^\alpha$ is taken from the experimental
RSCU values \cite{Duret}.
The synonymous mutability of amino acids is observed to be
higher in the short genes than in the long
genes for the nematode, Figure \ref{drosynon}. 
Indeed, of the amino acids, only arginine has a demonstrably lower
synonymous mutability for the short genes, as seen in 
Figure \ref{drosynon}.
We calculate the probability that the observed increase in
synonymous mutability is due to chance.
The probability of 17 or more amino acids showing this
trend out of 18 by chance is
$[
\left( \hbox{}_{18}^{18} \right)
+
\left( \hbox{}_{17}^{18} \right)
] 2^{-18} = 7.2 \times 10^{-5}$.
Making the same plot for the nematode, one observes the pattern to be
even more striking \cite{Tan,Nematode} (data not shown).
Indeed, of the amino acids, only proline has a demonstrably lower
synonymous mutability for the short genes, and only two other amino acids 
have roughly the same synonymous mutability in short and long genes.
The probability of 15 or more amino acids showing this
trend out of 16 by chance is
$[
\left( \hbox{}_{16}^{16} \right)
+
\left( \hbox{}_{15}^{16} \right)
] 2^{-16} = 2.6 \times 10^{-4}$.

While there are selective pressures on synonymous codon usage,
such as preference for tRNAs at different levels of abundance,
it seems unlikely that there would be a selection on the quantity
synonymous mutation rate, in and of itself, that is
significant enough to cause the observed correlation.
In other words, there are known to be selective pressures on codon usage.
What is not clear is why there should be selective pressure
on synonymous mutation rate itself.
There is selection pressure on the ability to adapt, however.
In order for short genes to evolve at an overall rate comparable
to that of long genes, the mutation rate per base 
would have to be higher
in short genes.
If one assumes that on average there is a certain number of mutations needed
to effect functional adaptation of a protein, and
that short proteins and long proteins need to evolve at roughly similar rates,
this then implies that short proteins need a higher per base rate of evolution
than long proteins---because they are shorter, and the evolution rate of a gene
is the evolution rate per base times the number of bases.  Thus, the evolution
rate per base must be higher for shorter proteins.
In contrast to Figure \ref{drosynon}, however, 
a correlation between conservative
or non-conservative mutation rate and gene length
was not observed for either \emph{Drosophila} or
the nematode (data not shown).

\subsection{Modulation of Recombination Rates}

An alternative means of evolution is  recombination, and
recombination rates are known to be correlated with 
codon usage bias \cite{Comeron1}.  
Selection pressure on short genes for greater evolvability could
favor a higher recombination rate per base, 
thus allowing short genes to evolve
at a rate comparable to that of long genes.  It would be unfavorable if
evolution for higher recombination rates led to lower conservative
or non-conservative mutation rates. 
 C+G content is known to be a rough measure of recombination rate 
\cite{Eyre-Walker2,Comeron4,Duret2000,Birdsell2002}.
In other words,
the correlation between C+G content and recombination rate is strong
enough that C+G content is now felt to be a useful maker of
local recombination rate \cite{Fullerton2001,Birdsell2002}.
Interestingly, we find that C+G is positively correlated with
all three mutation rates and is most highly correlated with
synonymous mutation rate. 
 Moreover, as Figure \ref{shortlong}a shows,
the codon usage of short genes is such that a higher per base rate
of estimated recombination is favored.  
The recombination rate of amino acid $\alpha$ is estimated by
\begin{equation}
{\rm estimated~recombination~rate}(\alpha) =
   \sum_{i \in \alpha} p_i^\alpha \times
({\rm number~of~C~or~G~bases~in~codon~} i) \ ,
\end{equation}
where the codon usage $p_i^\alpha$ is taken from the experimental
RSCU values \cite{Duret}.
In Figure \ref{shortlong}a, only one exception, for proline,
is found to the general pattern.
As Figure \ref{shortlong}b shows, a similar correlation between codon
usage and enhanced estimated recombination frequency is also observed 
in \emph{Drosophila}.  
No exceptions to the general pattern are found in Figure \ref{shortlong}b.
Finally, Figure \ref{shortlong}c shows the estimated recombination
rate for \emph{A.\ thaliana}.  In Figure \ref{shortlong}c,
only one exception, for glycine, is
found to the general pattern.
Considering all three species,
the probability of 52 or more amino acids showing this
trend out of 54 by chance is
$[
\left( \hbox{}_{52}^{54} \right)
+
\left( \hbox{}_{53}^{54} \right)
+
\left( \hbox{}_{54}^{54} \right)
] 2^{-54} = 8.2 \times 10^{-14}$.
The pattern is, thus, highly statistically significant.
One explanation for the
observed codon usage of short, high-expression genes is selective
pressure on crossover frequency.
On a long time scale, other factors such as neutral
evolution and rearrangements become important, and this is
likely the reason for the relatively modest shifts in the
codon usage observed in Figure \ref{shortlong}.

In
Figure \ref{recombine}a is shown the \emph{measured} recombination
rate versus protein length for genes in \emph{Drosophila}
at high expression levels \cite{Hey2002} (EST $>$ 50).  In this species,
codon bias is observable for genes at all recombination levels.
The correlation between codon bias and recombination rate is
seen, however, only when the latter is low
rates \cite{Hey2002,Marais2002}.  Figure \ref{recombine}
is, therefore, made only for
recombination rates less than 1 centimorgan per megabase.  A negative
correlation between recombination rate and protein length is observed.
In Figure \ref{recombine}c, the \emph{measured} recombination rate
versus protein length is shown for \emph{C.\ elegans} for genes
at high expression levels \cite{Marais2002}.
A clear negative
correlation between recombination rate and gene length is again observed.

\section{Discussion}


\subsection{Selective Pressures on Codon Mutation Rates}

It was found that for the Taq polymerase, nonpolar amino
acids are mutated at an elevated rate.
Nonpolar amino acids are more frequently present in the
interior cores of proteins, and mutations of these amino
acids more often lead to dramatic rearrangements of
the protein structure.
The pattern in the error-prone PCR mutation plot suggests that
the mutations that occur will tend to cause larger changes in
the structure of the encoded protein.
It is  becoming more accepted that
large mutation events such as transpositions, horizontal transfers, gene
exchange, and non-conservative mutations are necessary for dramatic
evolution.  This was shown quantitatively in \cite{Deem}.
  Non-conservative
mutations in the core of the protein would be one of the most
dramatic amino acid substitution moves possible
and can be considered to search the protein sequence space
most broadly.  In other words, under error-prone conditions,
 the \emph{Taq} polymerase favors codons for
the nonpolar amino acids that mutate non-conservatively.
This property of error-prone PCR greatly enhances the ability of this
method to improve protein function effectively by forcing
the search of greater regions of tertiary fold space. 
Moreover, the average mutational tendencies of \emph{Taq} 
can be modulated by codon usage.
Table \ref{oligocodon}
 defines codons by their tendencies to evolve under error
prone conditions. These data can be useful in the design of protein 
evolution experiments, especially when trying to evolve new motifs
\emph{ab initio}.

It was found that for V regions of mouse antibodies there is
an increase in the mutation rate of the charge amino acids.
These trends are not sensitive to whether 
equation \ref{1} or equation \ref{2a} is used to model
the mutation matrix or whether the mutation data are taken from
\cite{Smith,Shapiro} or from \cite{Milstein1995}.
Antibody V regions undergo DNA swapping of gene
fragments in order to create
the primary repertoire needed to develop resistance to disease.
Therefore, base mutations that
alter the framework of the proteins become less necessary.
More significant are mutations that lead to a greater binding
affinity.  
In protein-protein complexes, a positive correlation is observed
between binding affinity and the number of ionic interactions spanning
an interface \cite{Honig,Nussinov}. Thus for the 
polar amino acids participating in
binding, high conservative and non-conservative mutabilities would be
most favorable, since such characteristics would enable more efficient
searching of sequence space to optimize binding.

\subsection{Selective Pressures on Recombination Rates}


Previously, a correlation 
between codon usage bias and gene length 
had been observed in the species considered here \cite{Duret}.
Several mechanisms that might explain the increased 
codon bias in short genes were considered, including biased
tRNA levels, but all predicted
increased bias for longer genes, in contrast to the greater
observed bias for shorter genes \cite{Duret}.
We suggest that codon usage in short genes in these species has evolved
due to selection for increased recombination frequency, Figure
\ref{recombine}.  
This mechanism is consistent with previously observed
positive correlations between recombination rate and codon usage bias
and with previously observed negative correlations
between gene length and codon usage bias
\cite{Comeron1,Comeron4}.
The observed correlation between codon usage  
and synonymous mutation rate,
Figure \ref{drosynon}, may be a byproduct of selection
on recombination rate, as synonymous mutation
rate is positively correlated with C+G content
($R = 0.62$ 
 for
\emph{Drosophila}, and $R = 0.51$ for the nematode).

In \cite{Duret}, the codon usage bias was highest for those genes
at high expression levels, and Figure  \ref{shortlong} is based
upon those data.  In fact, the expression level was estimated
in \cite{Duret} from the frequency
with which those genes were observed in the
EST database.  It is possible that certain genes may 
be overrepresented in the EST database, in a way that is
correlated with the gene length.
If this unknown bias were the cause of the correlation in
Figure  \ref{shortlong}, then the opposite or no correlation would be
expected to be observed for genes at low expression.
In fact (data not shown), the same patterns observed
in Figure \ref{shortlong}a  are observed when codon usage for the 
genes at low (bottom 1/3 of genes with non-zero EST abundance)
rather than high (top 1/3 of genes with non-zero EST abundance)
expression levels are used:
Among the 54 amino acids, only three have lower estimated recombination
rates for the short genes at low expression levels
than for the long genes at low expression levels.

It might be argued that to be fully consistent with our theory,
the relevant recombination rate is that of the whole gene,
divided by the coding length of the gene.  This quantity is slightly different
from the quantity plotted in Figure \ref{recombine}ac because the
intron to exon composition of genes could vary systematically
with length.  This concern has been addressed in Figure \ref{recombine}b,
where recombination rate times gene length divided by coding length
has been plotted.  The same negative correlation between
recombination rate and gene length is again observed.

For our explanation to be consistent, it must be the case that
\emph{Drosophila} and \emph{C.\ elegans} are, in some sense,
mutationally starved.  The very existence of the
Hill-Robertson effect in these species \cite{Marais2002} implies
that this is the case, because it implies that point mutation is
insufficient to evolve linked genes, and that recombination is
necessary to break the linkage.  The existence of related
effects, such as interference selection \cite{Comeron2002},
provides additional evidence for the same reasons.
Finally, the fact that codon bias is observable only for genes
at low recombination rates in \emph{Drosophila},
less than 1 \cite{Marais2002} or 1.5
\cite{Hey2002} cM/Mb, provides additional indirect evidence that
the selective pressure to increase evolution rates is strongest where
evolution is the slowest.

\section{Conclusion}

Previous treatments of the evolutionary biology of codon usage
have largely ignored the possibility that codon usage could \emph{affect}
mutation or recombination rates and have primarily focused on using
codon usage as 
a measure of selection.  We here suggest that not only can codon usage
affect mutation and recombination rates but also codon
usage has been selected to enhance functional gene adaptation
within the context of the genetic code.  This line of
reasoning is in accord with strategies for optimized design of
experimental protein molecular evolution protocols, where speed of
evolution is an explicit goal \cite{Deem,Maranas2002}.

In Nature there are numerous examples of exploiting codon potentials
in ongoing evolutionary processes.  In the V regions of
encoded antibodies, high-potential serine codons such as AGC
are found predominately in the encoded CDR loops while the encoded
frameworks contain low-potential serine codons such as TCT 
\cite{Wagner}.
Unfortunately, antibodies and drugs are often no match for the
hydrophilic, high-potential codons of ``error-prone'' pathogens.
The dramatic mutability of the HIV gp120 coat protein is one
such example.  One can envision a
scheme for using codon potentials to target disease epitopes that
mutate rarely (\emph{i.e.}, low-potential) and unproductively
(\emph{i.e.}, become stop, low-potential, or structure-breaking
codons).  Such a therapeutic scheme should be
generally useful against diseases that use error prone
replication to escape  therapeutic treatments or vaccines.

\section{Acknowledgment}
This research was supported by the National Institutes of Health
and the National Science Foundation.


\newpage
\bibliography{codon}

\begin{thebibliography}{55}
\expandafter\ifx\csname natexlab\endcsname\relax\def\natexlab#1{#1}\fi

\bibitem[Altenberg, 1994]{Altenberg}
Altenberg L (1994) The evolution of evolvability in genetic programming.
\newblock In: Kinnear KE (ed.), \emph{Advances in Genetic Programming}. MIT
  Press, Cambridge, MA, pp. 47--74

\bibitem[Birdsell, 2002]{Birdsell2002}
Birdsell JA (2002) Integrating genomics, bioinformatics, and classical genetics
  to study to effects of recombination on genome evolution.
\newblock Mol Biol Evol 19:1181--1197

\bibitem[Blouin \emph{et~al.}, 1998]{Nematode}
Blouin MS, Yowell CA, Courtney CH, Dame JB (1998) Substitution bias, rapid
  saturation, and the use of {mtDNA} for nematode systematics.
\newblock Mol Biol Evol 15:1719--1727

\bibitem[Bogarad and Deem, 1999]{Deem}
Bogarad LD, Deem MW (1999) A hierarchical approach to protein molecular
  evolution.
\newblock Proc Natl Acad Sci USA 96:2591--2595

\bibitem[Bull \emph{et~al.}, 2001]{Bull2000}
Bull HJ, Lombardo MJ, Rosenberg SM (2001) Stationary-phase mutation in the
  bacterial chromosome: Recombination protein and {DNA} polymerase {IV}
  dependence.
\newblock Proc Natl Acad Sci USA 98:8334--8341

\bibitem[Comeron and Kreitman, 2000]{Comeron4}
Comeron JM, Kreitman M (2000) The correlation between intron length and
  recombination in \emph{Drosophila}: {D}ynamic equilibrium between mutational
  and selective forces.
\newblock Genetics 156:1175--1190

\bibitem[Comeron and Kreitman, 2002]{Comeron2002}
Comeron JM, Kreitman M (2002) Population, evolutionary and genomic consequences
  of interference selection.
\newblock Genetics 161:389--410

\bibitem[Comeron \emph{et~al.}, 1999]{Comeron1}
Comeron JM, Kreitman M, Aguade M (1999) Natural selection on synonymous sites
  is correlated with gene length and recombination in \emph{Drosophila}.
\newblock Genetics 151:239--249

\bibitem[Dayhoff \emph{et~al.}, 1978]{PAM}
Dayhoff MO, Schwartz RM, Orcutt BC (1978) A model of evolutionary change in
  proteins.
\newblock In: \emph{Atlas of Protein Sequence and Structure}, National
  Biomedical Research Foundation, vol.~5, pp. 345--352

\bibitem[Durbin \emph{et~al.}, 1998]{Durbin}
Durbin R, Eddy S, Krogh A, Mitchison G (1998) Biological Sequence Analysis.
\newblock Cambridge University Press, Cambridge, United Kingdom

\bibitem[Duret \emph{et~al.}, 2000]{Duret2000}
Duret L, Marais G, Bi{\'e}mont C (2000) Transposons but not retrotransposons
  are located preferentially in regions of high recombination rate in \emph{C.
  elegans}.
\newblock Genetics 156:1661--1669

\bibitem[Duret and Mouchiroud, 1999]{Duret}
Duret L, Mouchiroud D (1999) Expression pattern and, surprisingly, gene length
  shape codon usage in \emph{Caenorhabditis}, \emph{Drosophila}, and
  \emph{Arabidopsis}.
\newblock Proc Natl Acad Sci USA 96:4482--4487

\bibitem[Epstein, 1966]{Epstein}
Epstein CJ (1966) Role of the amino acid `code' and of selection for
  conformmation in the evolution of proteins.
\newblock Nature 210:25--28

\bibitem[Eyre-Walker, 1993]{Eyre-Walker2}
Eyre-Walker A (1993) Recombination and mammalian genome evolution.
\newblock Proc R Soc Lond Ser B Biol Sci 252:237--243

\bibitem[Fitch, 1966]{Fitch}
Fitch WM (1966) The relation between frequencies of amino acids and ordered
  trinucleotides.
\newblock J Mol Biol 16:1--16

\bibitem[Freire, 2002]{Freire2002}
Freire E (2002) Designing drugs against heterogeneous targets.
\newblock Nat Biotech 20:15--16

\bibitem[Friedberg \emph{et~al.}, 2000]{Friedberg2000}
Friedberg EC, Feaver WJ, Gerlach VL (2000) The many faces of {DNA} polymeraces:
  {S}trategies for mutagenesis and for mutational avoidance.
\newblock Proc Natl Acad Sci USA 97:5681--5683

\bibitem[Fullerton \emph{et~al.}, 2001]{Fullerton2001}
Fullerton SM, Carvalho AB, Clark AG (2001) Local rates of recombination are
  positively correlated with {GC} content in the human genome.
\newblock Mol Biol Evol 18:1139--1142

\bibitem[Goldberg and Wittes, 1966]{Goldberg}
Goldberg AL, Wittes R (1966) Genetic code: {A}spects of organization.
\newblock Science 153:420--422

\bibitem[Goldman and Yang, 1994]{Yang2}
Goldman N, Yang Z (1994) A codon-based model of nucleotide substitution for
  protein-coding {DNA} sequences.
\newblock Mol Biol Evol 17:32--43

\bibitem[Henikoff and Henikoff, 1992]{BLOSUM}
Henikoff JG, Henikoff S (1992) Amino acid substitution matrices from protein
  blocks.
\newblock Proc Natl Acad Sci USA 89:10915--10919

\bibitem[Hey and Kliman, 2002]{Hey2002}
Hey J, Kliman RM (2002) Interactions between natural selection, recombination,
  and gene density in the genes of {D}rosophila.
\newblock Genetics 160:595--608

\bibitem[Kepler, 1997]{Kepler1997}
Kepler TB (1997) Codon bias and plasticity in immunoglobulins.
\newblock Mol Biol Evol 14:637--643

\bibitem[Lathrop and Pazzani, 1999]{Lathrop99}
Lathrop RH, Pazzani MJ (1999) Combinatorial optimization in rapidly mutating
  drug-resistant viruses.
\newblock J Comb Optim 3:301--320

\bibitem[Lawrence, 1997]{Lawrence1997}
Lawrence JG (1997) Selfish operons and speciation by gene transfer.
\newblock Trends Microbiol 5:355--359

\bibitem[Li \emph{et~al.}, 1985]{LiWuLuo}
Li W, Wu C, Luo C (1985) A new method for estimating synonymous and
  nonsynonymous rates of nucleotide substitution considering the relative
  likelihood of nucleotide and codon changes.
\newblock Mol Biol Evol 2:150--174

\bibitem[Lutz and Benkovic, 2000]{Benkovic2000}
Lutz S, Benkovic SJ (2000) Homology-independent protein engineering.
\newblock Curr Opin Biotech 11:319--324

\bibitem[Marais and Piganeau, 2002]{Marais2002}
Marais G, Piganeau G (2002) Hill-{R}obertson interference is a minor
  determinant of variations in codon bias across \emph{Drosophilia
  Melanogaster} and \emph{Caenorhabditis elegans} genomes.
\newblock Mol Biol Evol 19:1399--1406

\bibitem[Moore and Maranas, 2000]{Maranas}
Moore GL, Maranas CD (2000) Modeling {DNA} mutation and recombination for
  directed evolution experiments.
\newblock J Theor Biol 205:483--503

\bibitem[Moore and Maranas, 2002]{Maranas2002}
Moore GL, Maranas CD (2002) {eCodonOpt}: A systematic computational framework
  for optimizing codon usage in directed evolution experiments.
\newblock Nucleic Acids Res 30:2407--2416

\bibitem[Neuberger and Milstein, 1995]{Milstein1995}
Neuberger MS, Milstein C (1995) Somatic hypermutation.
\newblock Curr Opin Immunol 7:248--254

\bibitem[Ofria \emph{et~al.}, 1999]{Ofria1999}
Ofria C, Adami C, Collier TC, Hsu GK (1999) Evolution of differentiated
  expression patterns in digital organisms.
\newblock Adv Artific Life 1674:129--138

\bibitem[Patel \emph{et~al.}, 2001]{Patel2001}
Patel PH, Kawate H, Adman E, Ashbach M, Loeb LA (2001) A single highly mutable
  catalytic site amino acid is critical for {DNA} polymerase fidelity.
\newblock J Biol Chem 276:5044--5051

\bibitem[Patten \emph{et~al.}, 1997]{Stemmer1997}
Patten PA, Howard RJ, Stemmer WPC (1997) Applications of {DNA} shuffling to
  pharmaceuticals and vaccines.
\newblock Curr Opin Biotech 8:724--733

\bibitem[Pennisi, 1998]{Pennisi1998}
Pennisi E (1998) Molecular evolution--{H}ow the genome readies itself for
  evolution.
\newblock Science 281:1131--1134

\bibitem[Peper, 2003]{Pepper}
Peper JW (2003) The evolution of evolvability in genetic linkage patterns.
\newblock Biosystems 69:115--126

\bibitem[Petrounia and Arnold, 2000]{Arnold2000}
Petrounia IP, Arnold FH (2000) Designed evolution of enzymatic properties.
\newblock Curr Opin Biotech 11:325--330

\bibitem[Petrov and Hartl, 1999]{Petrov}
Petrov DA, Hartl DL (1999) Patterns of nucleotide substitution in
  \emph{Drosophila} and mammalian genomes.
\newblock Proc Natl Acad Sci USA 96:1475--1479

\bibitem[Plotkin and Doshoff, 2003]{Plotkin2003}
Plotkin JB, Doshoff J (2003) Codon bias and frequency-dependent selection on
  the hemagglutinin epitopes of influenza {A} virus.
\newblock Proc Natl Acad Sci USA 100:7152--7157

\bibitem[Shapiro \emph{et~al.}, 1999]{Shapiro}
Shapiro GS, Aviszus K, Ikle D, Wysocki LJ (1999) Predicting regional mutability
  in antibody {V} genes based solely on di- and trinucleotide sequence
  composition.
\newblock J Immunol 163:259--268

\bibitem[Sharp \emph{et~al.}, 1986]{Sharp1986}
Sharp PM, Tuohy TMF, Mosurski KR (1986) Codon usage in yeast: {C}luster
  analysis clearly differentiates highly and lowly expressed genes.
\newblock Nucleic Acids Res 14:5125--5143

\bibitem[Sheinerman \emph{et~al.}, 2000]{Honig}
Sheinerman FB, Norel R, Honig B (2000) Electrostatic aspects of protein-protein
  interactions.
\newblock Curr Opin Struct Biol 10:153--159

\bibitem[Shen \emph{et~al.}, 2000]{Storb2000}
Shen HM, Michael N, Kim N, Storb U (2000) The {TATA} binding protein, c-{M}yc
  and survivin genes are not somatically hypermutated, while {I}g and {BCL}6
  genes are hypermutated in human memory {B} cells.
\newblock Int Immun 12:1085--1093

\bibitem[Smith \emph{et~al.}, 1996]{Smith}
Smith DS, Creadon G, Jena PK, Portanova JP, Kotzin BL, Wysocki LJ (1996) Di-
  and trinucleotide target preference of somatic mutagenesis in normal and
  autoreactive {B} cells.
\newblock J Immunol 156:2642--2652

\bibitem[Storb, 2001]{Storb2001}
Storb U (2001) {DNA} polymerases in immunity: {P}rofiting from errors.
\newblock Nat Immunol 2:484--485

\bibitem[Sutton and Walker, 2001]{Walker2000}
Sutton MD, Walker GC (2001) Managing {DNA} polymerases: {C}oordinating {DNA}
  replication, {DNA} repair, and {DNA} recombination.
\newblock Proc Natl Acad Sci USA 98:8342--8349

\bibitem[Tan, 2002]{Tan}
Tan T (2002) Ph.D. Dissertation. The Codon Mutation Matrix in the Context of
  Protein Molecular Evolution.
\newblock UCLA

\bibitem[Thearling and Ray, 1997]{Ray1997}
Thearling K, Ray TS (1997) Evolving parallel computation.
\newblock Complex Systems 10:229--237

\bibitem[Travis and Travis, 2002]{Travis}
Travis JMJ, Travis ER (2002) Mutator dynamics in fluctuating environments.
\newblock Proc Roy Soc B London 269:591--597

\bibitem[Volkenstein, 1994]{Volkenstein}
Volkenstein MV (1994) Physical Approaches to Biological Evolution.
\newblock Springer-Verlag, New York

\bibitem[Wagner and Altenberg, 1996]{Wagner1996}
Wagner GP, Altenberg L (1996) Perspective: Complex adaptations and the
  evolution of evolvability.
\newblock Evolution 50:967--976

\bibitem[Wagner \emph{et~al.}, 1995]{Wagner}
Wagner SD, Milstein C, Neuberger MS (1995) Codon bias targets mutation.
\newblock Nature 376:732

\bibitem[Woese, 1965]{Woese}
Woese CR (1965) On the evolution of the genetic code.
\newblock Proc Natl Acad Sci USA 54:1546--1552

\bibitem[Xu \emph{et~al.}, 1997]{Nussinov}
Xu D, Lin SL, Nussinov R (1997) Protein binding versus protein folding: {T}he
  role of hydrophobic bridges in protein association.
\newblock J Mol Biol 265:68--84

\bibitem[Yang and Kumar, 1996]{Yang}
Yang Z, Kumar S (1996) Approximate methods for estimating the pattern of
  nucleotide substitution and the variation of substitution rates among sites.
\newblock Mol Biol Evol 13:650--659

\end{thebibliography}
\clearpage

\renewcommand{\baselinestretch}{1.0} \tiny\normalsize

\newpage
\begin{table}
\centering
\begin{minipage}[t]{6.00in}
\centering
\renewcommand{\footnoterule}{}
\caption{Table of codon classifications for the error-prone PCR
system.}
\label{oligocodon}
\begin{tabular}{p{1.3in}p{1.3in}p{1.3in}p{1.3in}}
\hline
&\multicolumn{1}{l}{synonymous}&\multicolumn{1}{l}{conservative}&\multicolumn{1}{l}{non-conservative}\\
\hline
Cys&   &   &TGC, TCT\\
Ser&TCA, TCT&TCC, TCG&AGC, AGT\\
Thr&ACA&ACC, ACG, ACT&   \\
Pro&CCA, CCC, CCT&CCG&   \\
Ala&GCA, GCT&GCC, GCG&   \\
Gly&GGA, GGT&   &GGC, GGG\\
Asn&   &   &AAC, AAT\\
Gln&   &   &CAA, CAG\\
Asp&   &   &GAC, GAT\\
Glu&   &   &GAA, GAG\\
His&   &   &CAC, CAT\\
Arg&CGA, CGT&   &AGA, AGG, CGC\\
   &   &   &CGG\\
Lys&   &   &AAA, AAG\\
Met&   &ATG&   \\
Ile&   &ATA&ATC, ATT\\
Leu&CTA&   &CTC, CTG, CTT\\
   &   &   &TTA, TTG\\
Val&   &   &GTA, GTC, GTG\\
   &   &   &GTT\\
Phe&   &   &TTC, TTT\\
Tyr&   &   &TAC, TAT\\
Trp&   &   &TGG\\
Stop&   &   &TAA, TAG, TGA\\
\end{tabular}
\end{minipage}
\end{table}
\clearpage

\clearpage

\newpage
\thispagestyle{empty}
\clearpage

\newpage
\thispagestyle{empty}
\clearpage

\newpage

\newpage

\newpage
\thispagestyle{empty}
\begin{figure}[htbp]
\centering
\leavevmode
\psfig{file=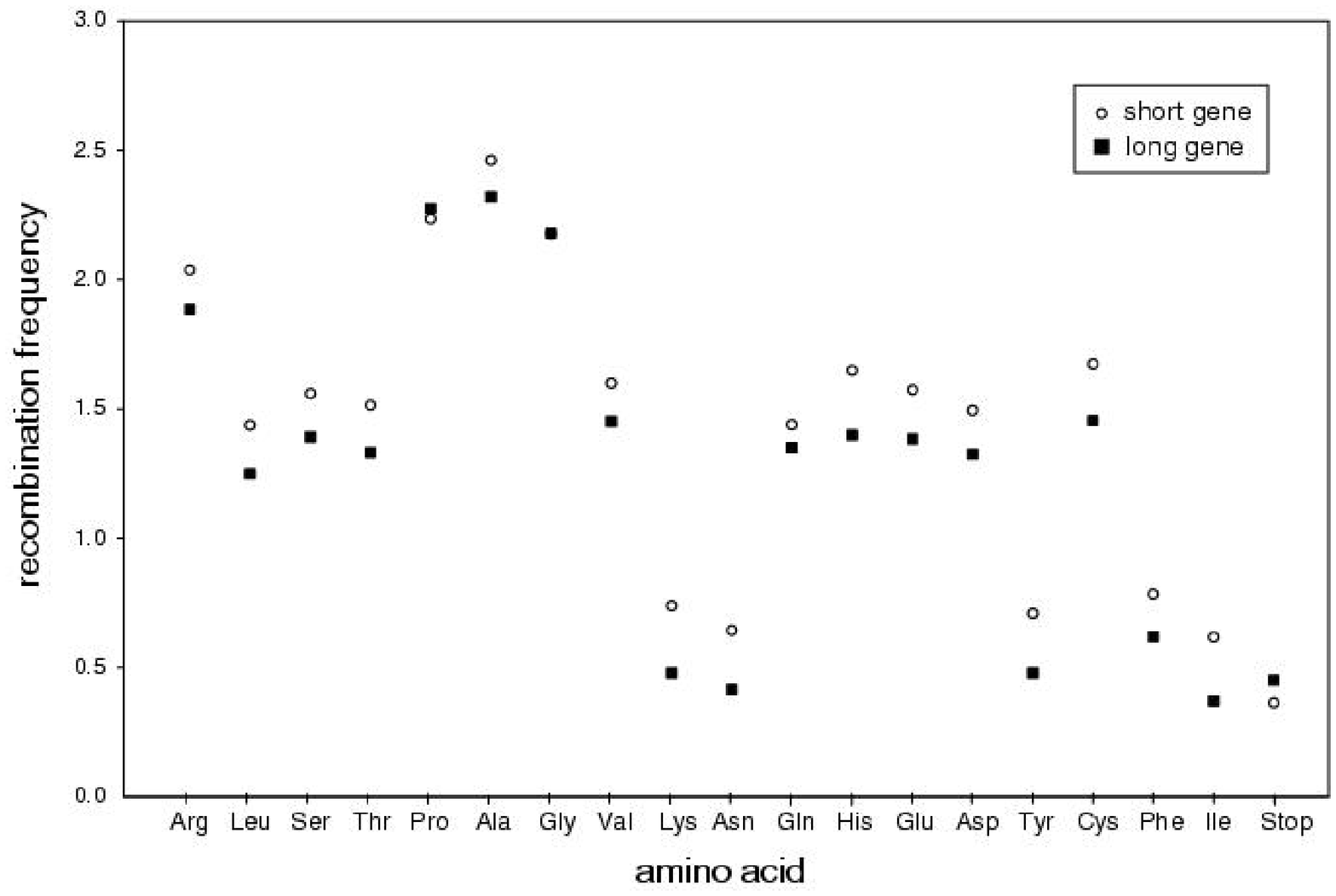,height=8cm}~\\
\centering
\leavevmode
\psfig{file=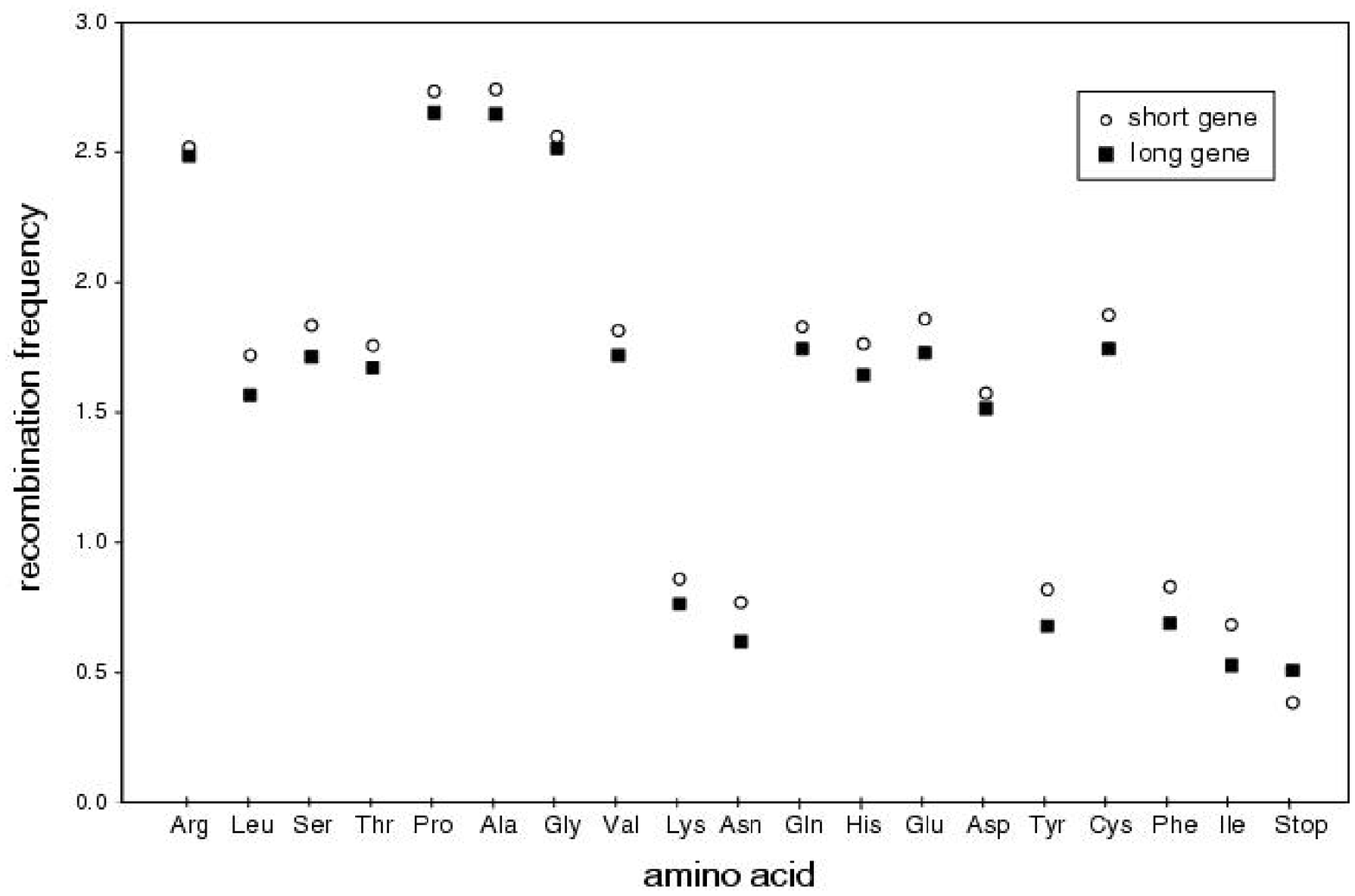,height=8cm}~\\
\centering
\leavevmode
\psfig{file=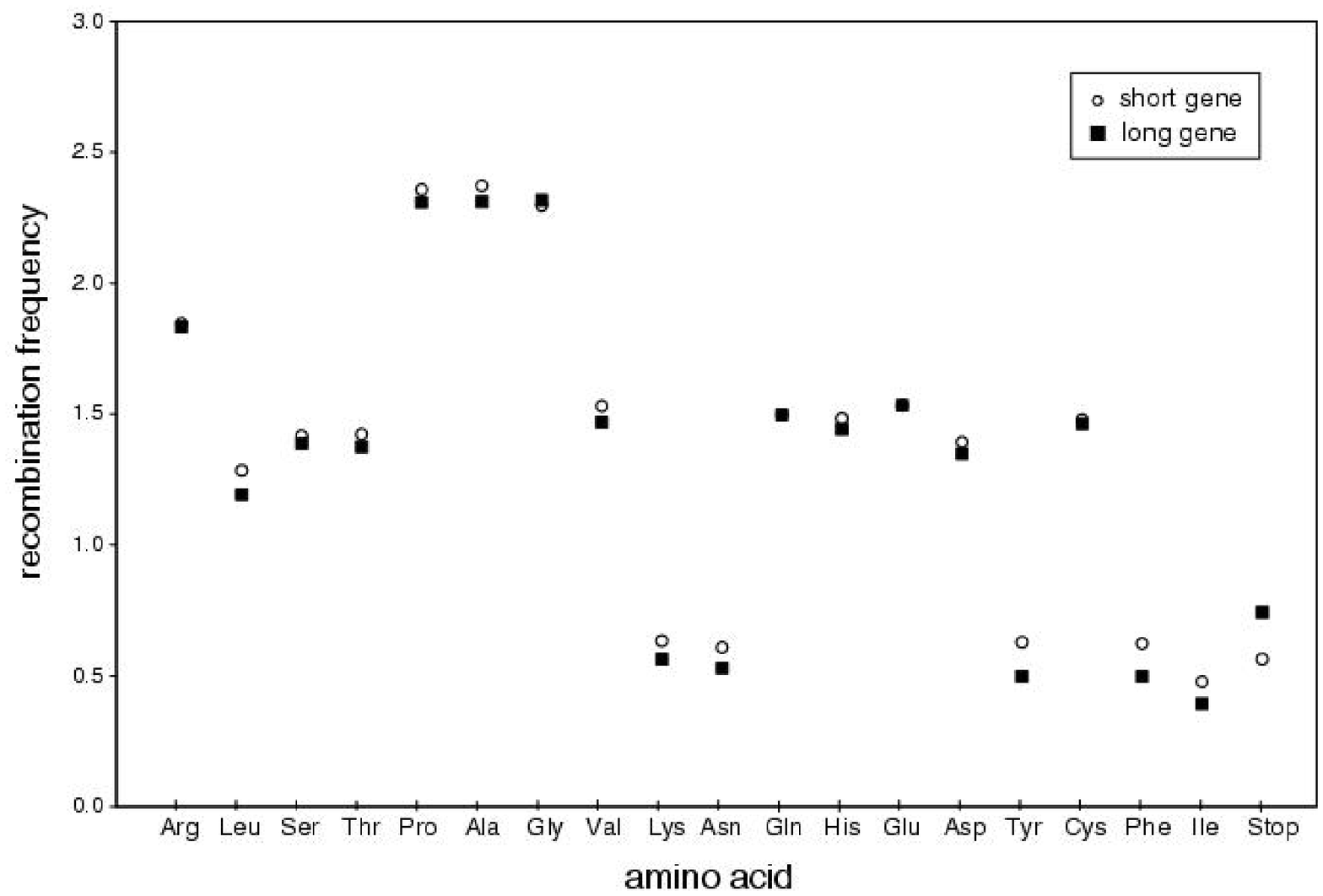,height=8cm}
\end{figure}
\setcounter{figure}{4}
\begin{figure}[htbp]
\caption{Estimated recombination frequency for a) \emph{C.\ elegans},
b) \emph{D.\ melanogaster}, and
c) \emph{A.\ thaliana}
for amino acids in short ($< 333$ amino acids)
and long ($> 570$ amino acids) genes at high expression levels
(top 1/3 of genes with non-zero EST abundance).
Higher values of estimated
recombination frequency are observed in the shorter genes.
Recombination frequency is estimated by the sum over all codons
encoding a given amino acid of the observed codon usage times the number of
C and G bases in the codon.}
\label{shortlong}
\end{figure}

\clearpage
\thispagestyle{empty}
\begin{figure}[htbp]
\centering
\leavevmode
\psfig{file=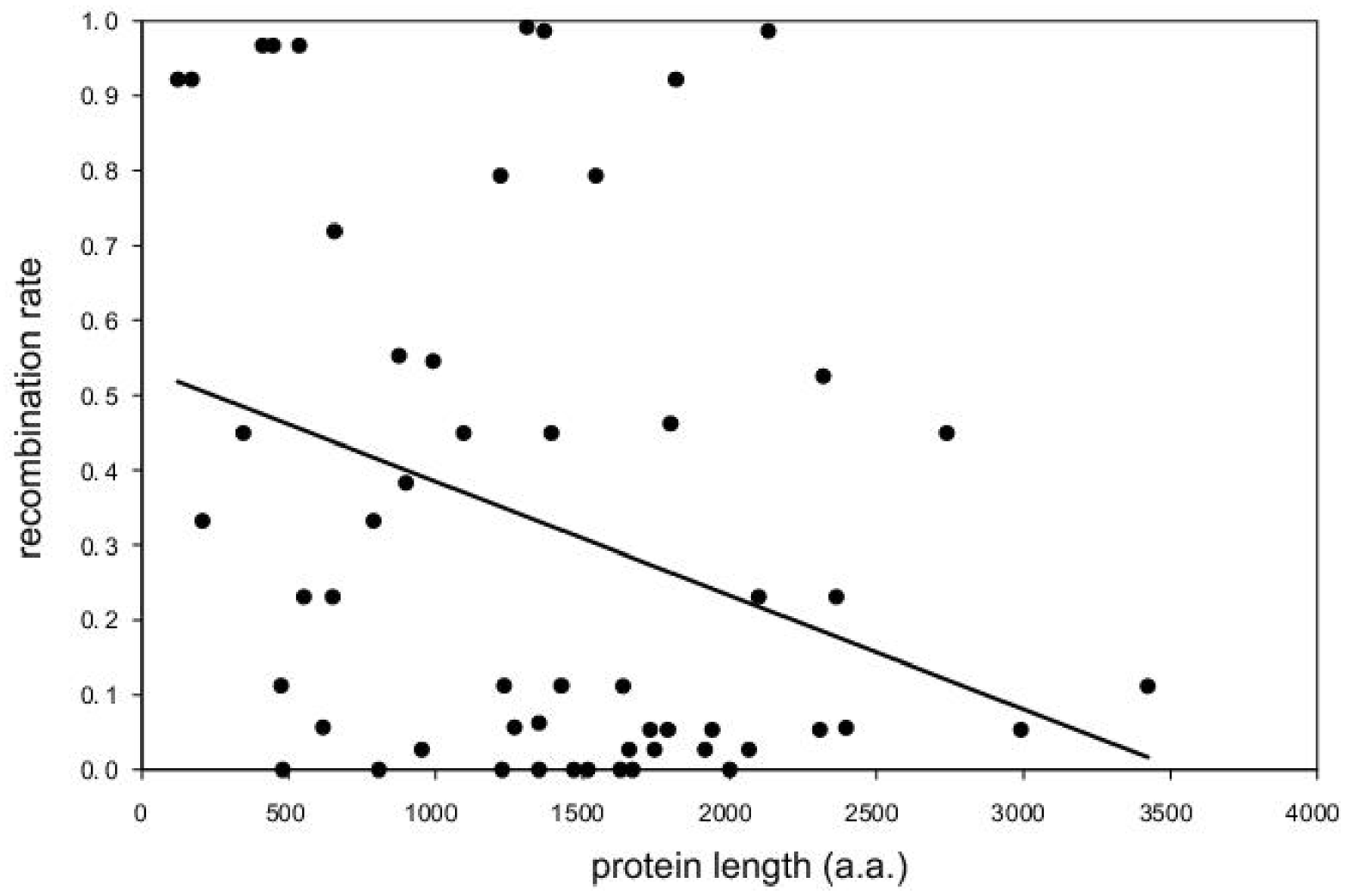,height=8cm}~\\
\centering
\leavevmode
\psfig{file=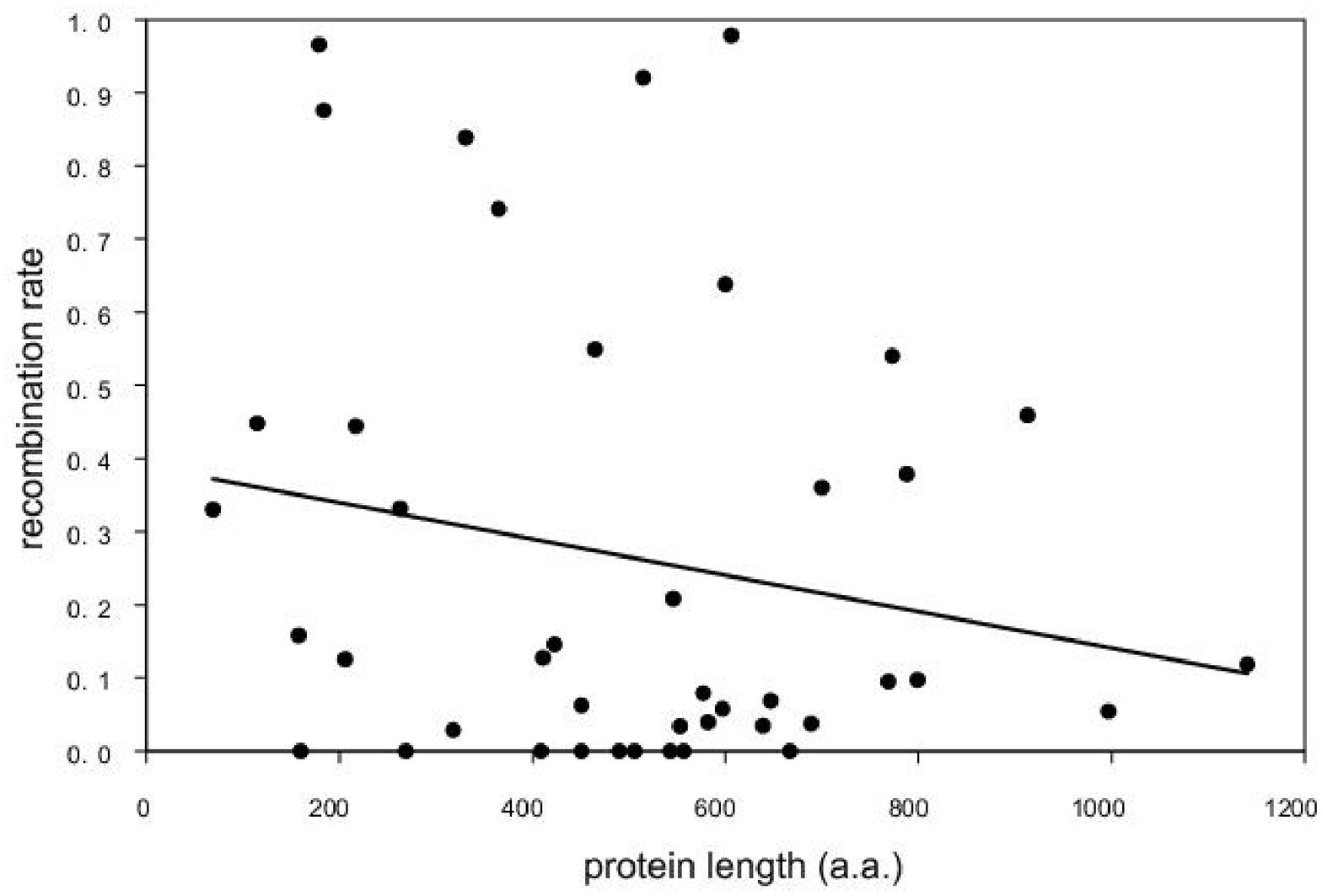,height=8cm}~\\
\centering
\leavevmode
\psfig{file=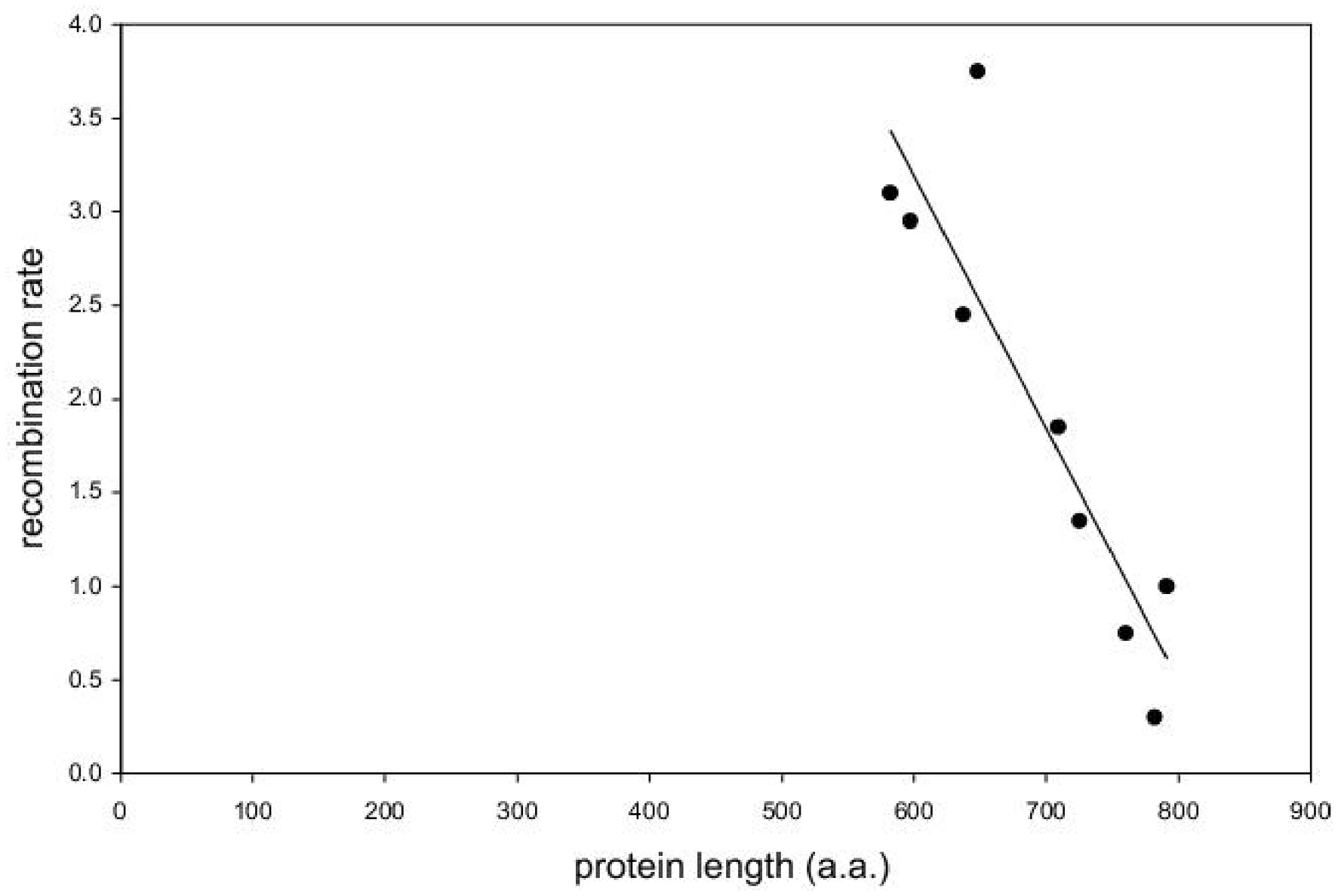,height=8cm}
\end{figure}
\setcounter{figure}{5}
\begin{figure}[htbp]
\caption{
   Measured recombination frequency (centimorgan/megabase) as a 
   function of protein length (amino acids) for 
a) \emph{D.\ melanogaster},
b) \emph{D.\ melanogaster}, where recombination frequency is
   modifed to account for intron to exon
   base composition, $R \times ({\rm gene~length}) / ({\rm coding~length})$,
   and
c) \emph{C.\ elegans}.
  Also shown are linear fits to the data; the correlation coefficients
  are a) $R = -0.32$, b) $R = -0.20$, and c) $R = -0.89$.
  All data are for genes at high expression levels.
  Data in a) and b) are taken from \cite{Hey2002}.
  Data in c) are replotted from the binned data of \cite{Marais2002}.
}
\label{recombine}
\end{figure}

\clearpage

\begin{center}
{\bf Supplementary Information}
\end{center}

\begin{table}[h]
\begin{minipage}[h]{6.00in}
\caption{The base mutation matrix $t$ for error-prone PCR \cite{Maranas}.}
\label{tmatrixpcr}
\begin{tabular}{rrrrr}
\hline
 &\multicolumn{1}{l}{A}&\multicolumn{1}{l}{C}&\multicolumn{1}{l}{G}&\multicolumn{1}{l}{T}\\
\hline
\\
A&1--7.4$\times 10^{-6}$&7.0$\times 10^{-7}$&3.2$\times
 10^{-6}$&3.5$\times 10^{-6}$\\
\\
C&7.0$\times 10^{-7}$&1--2.8$\times 10^{-6}$&0&2.1$\times 10^{-6}$\\
\\
G&2.1$\times 10^{-6}$&0&1--2.8$\times 10^{-6}$&7.0$\times 10^{-7}$\\
\\
T&3.5$\times 10^{-6}$&3.2$\times 10^{-6}$&7.0$\times 10^{-7}$&1--7.4$\times 10^{-6}$\\
\end{tabular}
\end{minipage}
\end{table}

\begin{table}[h]
\begin{minipage}[h]{6.00in}
\caption{The base mutation matrix $t$ for V regions of mouse antibodies
\cite{Smith,Shapiro}.$^a$
}
\label{tmatrixsmi}
\begin{tabular}{rrrrr}
\hline
 &\multicolumn{1}{l}{A}&\multicolumn{1}{l}{C}&\multicolumn{1}{l}{G}&\multicolumn{1}{l}{T}\\
\hline
\\
A&1--3.327$\times 10^{-5}$&6.550$\times 10^{-6}$&1.847$\times
 10^{-5}$&8.250$\times 10^{-6}$\\
\\
C&4.580$\times 10^{-6}$&1--2.597$\times 10^{-5}$&4.370$\times
 10^{-6}$&1.702$\times 10^{-5}$\\
\\
G&1.329$\times 10^{-5}$&6.200$\times 10^{-6}$&1--2.407$\times
 10^{-5}$&4.580$\times 10^{-6}$\\
\\
T&4.360$\times 10^{-6}$&7.270$\times 10^{-6}$&3.960$\times 10^{-6}$&1--1.559$\times 10^{-5}$\\
\end{tabular}
\end{minipage}
\end{table}
 \hbox{}$^a$For example $t_{AC} = 6.550\times 10^{-6}$.


\begin{table}[h]
\begin{minipage}[h]{6.00in}
\renewcommand{\footnoterule}{}
\caption{The base mutation matrix $t$ for \emph{Drosophila} \cite{Petrov}.
Only the relative rates are significant.}
\label{tmatrixdro}
\begin{tabular}{rrrrr}
\hline
 &\multicolumn{1}{l}{A}&\multicolumn{1}{l}{C}&\multicolumn{1}{l}{G}&\multicolumn{1}{l}{T}\\
\hline
\\
A&1--4.0$\times 10^{-5}$&1.0$\times 10^{-5}$&1.6$\times
 10^{-5}$&1.4$\times 10^{-5}$\\
\\
C&1.8$\times 10^{-5}$&1--6.0$\times 10^{-5}$&1.2$\times
 10^{-5}$&3.0$\times 10^{-5}$\\
\\
G&3.0$\times 10^{-5}$&1.2$\times 10^{-5}$&1--6.0$\times
 10^{-5}$&1.8$\times 10^{-5}$\\
\\
T&1.4$\times 10^{-5}$&1.6$\times 10^{-5}$&1.0$\times 10^{-5}$&1--4.0$\times 10^{-5}$\\
\end{tabular}
\end{minipage}
\end{table}

\begin{table}[h]
\begin{minipage}[h]{6.00in}
\caption{The base mutation matrix for human, non-immune system B-cell genes
\cite{Storb2000}.  Only the relative rates are significant.  Data were
generated from counts of observed mutations by the summing the
formula $N_{\rm observed~mutations~ x \to y}
= N_{\rm clones} N_{\rm bases} P(x) P(x \to y \vert x) $ over all genes
examined, and averaging the mutation matrix so obtained over the two donors.}
\label{tmatrixhuman}
\begin{tabular}{rrrrr}
\hline
 &\multicolumn{1}{l}{A}&\multicolumn{1}{l}{C}&\multicolumn{1}{l}{G}&\multicolumn{1}{l}{T}\\
\hline
\\
A&1--1.86$\times 10^{-5}$&0.0$\times 10^{-5}$&1.4$\times
 10^{-5}$&0.46$\times 10^{-5}$\\
\\
C&2.2$\times 10^{-5}$&1--5.73$\times 10^{-5}$&0.33$\times
 10^{-5}$&3.2$\times 10^{-5}$\\
\\
G&5.9$\times 10^{-5}$&0.0$\times 10^{-5}$&1--7.5$\times
 10^{-5}$&1.6$\times 10^{-5}$\\
\\
T&0.71$\times 10^{-5}$&2.3$\times 10^{-5}$&0.71$\times 10^{-5}$&1--3.72$\times 10^{-5}$\\
\end{tabular}
\end{minipage}
\end{table}

\begin{table}[h]
\begin{minipage}[h]{6.00in}
\caption{The base mutation matrix $t$ for mitochondrial DNA in a
nematode \cite{Nematode}. }
\label{tmatrixnem}
\begin{tabular}{rrrrr}
\hline
 &\multicolumn{1}{l}{A}&\multicolumn{1}{l}{C}&\multicolumn{1}{l}{G}&\multicolumn{1}{l}{T}\\
\hline
\\
A&1--7.60$\times 10^{-7}$&4.00$\times 10^{-8}$&6.90$\times
 10^{-7}$&3.00$\times 10^{-8}$\\
\\
C&2.20$\times 10^{-7}$&1--2.41$\times 10^{-6}$&1.00$\times
 10^{-7}$&2.09$\times 10^{-6}$\\
\\
G&3.02$\times 10^{-6}$&8.00$\times 10^{-8}$&1--3.15$\times
 10^{-6}$&5.00$\times 10^{-8}$\\
\\
T&3.00$\times 10^{-8}$&4.00$\times 10^{-7}$&1.00$\times 10^{-8}$&1--4.40$\times 10^{-7}$\\
\end{tabular}
\end{minipage}
\end{table}

\clearpage
\thispagestyle{empty}
\begin{center}
{\bf Caption for Cover Figure}
\end{center}

\bigskip
\bigskip

Emerging patterns from the inherent structure of the genetic code
and non-uniform mutation rates in error prone replication. The relative
rate of codon mutation above baseline (blue) is shown by color intensity.
Non-baseline synonymous changes are green; conservative, orange; and
non-conservative, red.  The codons are ordered by AAX, CAX, GAX, TAX, ACX,
\ldots.

\end{document}